\documentclass{PoS}

\usepackage{color}
\usepackage{amsmath,amssymb,amsfonts,wasysym}

\newcommand{\luna}{{\tt LUNA}}
\newcommand{\multi}{{\sc MultiNest}}
\newcommand{\nature}{Nature}
\newcommand{\science}{Science}
\newcommand{\apj}{ApJ}
\newcommand{\aanda}{A\&A}
\newcommand{\aandas}{A\&AS}
\newcommand{\aj}{AJ}
\newcommand{\mnras}{MNRAS}
\newcommand{\pnas}{PNAS}
\newcommand{\abio}{Astrobiology}
\newcommand{\icarus}{Icarus}
\newcommand{\pasj}{PASJ}

\title{In Search of Exomoons}

\ShortTitle{In Search of Exomoons}

\author{\speaker{David M. Kipping}\thanks{NASA Carl Sagan Fellow.}\\
        Harvard-Smithsonian Center for Astrophysics, Cambridge, MA, USA\\
        E-mail: \email{dkipping@cfa.harvard.edu}}

\abstract{
Two decades ago, astronomers began detecting planets orbiting stars other than 
our Sun, so-called exoplanets. Since that time, the rate of detections and the 
sensitivity to ever-smaller planets has improved dramatically with several 
Earth-sized planets now known. As our sensitivity dives into the terrestrial 
regime, increasingly the community has wondered if the moons of exoplanets may 
also be detectable, so-called ``exomoons''. Their detection represents an 
outstanding challenge in modern astronomy and would provide deep insights into 
the uniqueness of our Solar System and perhaps even expand the definition of 
habitability. Here, I will briefly review theoretical studies exploring the 
formation and evolution of exomoons, which serve to guide observational 
searches and provide testable hypotheses. Next, I will outline the different 
methods which have been proposed to accomplish this challenging feat and their 
respective merits. Finally, initial results from observational efforts will be 
summarized with a view to future prospects as well.
}

\FullConference{Frank N. Bash Symposium 2013: New Horizons in Astronomy (BASH 2013)\\
		October 6-8, 2013\\
		Austin, Texas }

\begin{document}

\section{INTRODUCTION} 

Within the field of astronomy, there are few areas of research which have 
enjoyed as much public enthusiasm and rapid rates of discovery than that
of exoplanetary science - a research area which was not even established until 
two decades ago. Astronomers first waded out around the shores of this novel
discipline around 1990, with compelling evidence for exoplanets found by
\cite{campbell:1988}, \cite{latham:1989} and 
\cite{wolszczan:1992}. However, it was not until the mid-nineties, 
after the discovery of 51 Pegasi b by \cite{mayor:1995}, that 
astronomers unfurled their sails and began regularly ensnaring impressive hauls 
of exoplanets.

Since then, more than one thousand confirmed exoplanets and five
thousand candidates have been detected. Furthermore, observational sensitivity 
has dramatically improved from being limited to hot-Jupiters back in the 
mid-nineties \cite{marcy:1998} to now being able to detect 
Earth-sized habitable-zone worlds \cite{batalha:2013}. This progress 
stems from a wide range of improvements in the instrumentation, 
telescope time, use of dedicated missions, man-power, observing strategy, 
modeling and of course research funding, which enables all of the above 
\cite{marcy:2014}.

This journey of scientific discovery is surely not complete though, and many
mysteries and outstanding challenges remain. For example, we now know that our 
solar system and our planet represent just a drop in the cosmic ocean, and the 
surprising diversity of planetary architectures and compositions indicates that 
we lack the deep understanding of how planets form and evolve 
\cite{youdin:2013}. Similarly, whilst empirical estimates of the occurence rate 
of habitable-zone rocky planets have been made \cite{fressin:2013,
petigura:2013}, the detection of bio-signatures or detailed 
atmospheric characterization of such worlds is still some way off 
\cite{rodler:2014}.

With our sensivity to exoplanets reaching the level of Earth and even sub-Earth
sized worlds \cite{batalha:2013}, one outstanding question which may 
be within reach is the detection of extrasolar moons; so-called ``exomoons''. 
Their detection would likely usher in an entire new sub-field of scientific 
endeavor, much like the wind change instigated by the discovery of 51-Pegasi-b. 
Measuring the physical and orbital characteristics of ensemble populations of 
exomoons would likely yield powerful constraints on the formation and evolution 
mechanisms guiding alien worlds \cite{gong:2013}. Such bodies may
themselves be habitable \cite{williams:1997,heller:2012a,heller:2012b,
forgan:2013} or affect the 
habitability of the planets which they orbit \cite{laskar:1993}. A 
confirmed exomoon detection would therefore represent a major scientific 
breakthrough, yet this lofty goal remains a daunting and outstanding challenge 
to modern astronomy.

In this work, I will briefly discuss several plausible formation mechanisms
for exomoons in \S2, with a view as to how these ideas may impact observational 
strategies. In \S3, I will review the proposed detection techniques and 
sensitivities, with focus on the transit-based methods. In \S5, empirical 
results to date will be summarized and placed into context. Finally, in \S6, we 
discuss future prospects for exomoon detection with current and planned
exoplanet missions.

\clearpage

\section{FORMATION \& EVOLUTION}

\subsection{Context}

We begin by briefly discussing several possible formation and subsequent 
evolution mechanisms for exomoons. Since we currently know of no confirmed 
exomoons, this topic is strongly influenced by inspection of the moons within 
our solar system, so-called ``endomoons''. Given the enormous diversity of
exoplanetary architectures discovered to date, which rarely resemble the
Solar System, it is important to not put too much faith in such theories.
Nevertheless, they provide a useful starting point and at least offer hypotheses
which can be subsequently tested with observations. I therefore endeavor to
provide only a brief introduction to give a flavor of how theory could guide
observing strategy.

In our solar system, we observe two broad classes of satellites, although there
exists no widely accepted definition as to what divides these two classes.
In this work, I define regular satellites to have formed in-situ from the
circumplanetary disk and these tend to display nearly-coplanar and 
nearly-circular orbits. In contrast, irregular satellites are those which formed 
via some other mechanism, such as an impact or a capture, and these tend to
exhibit highly inclined (including retrograde motion) and highly eccentric
orbital paths. Regular satellites also tend to have more compact orbits than
the irregular population. Naturally, these two populations have distinct
histories and thus I will split the discussion of formation and evolution 
accordingly.

\subsection{Formation of Regular Satellites}

Although considerable diversity exists within the literature for proposed
formation mechanisms of the regular satellites, the fact that such satellites 
have compact and prograde orbits has led to the general consensus that they 
formed from a circumplanetary disk \cite{mosqueira:2003a}. In 
general, formation models focus on the Jovian (belonging to Jupiter) and Kronian 
(belonging to Saturn) satellites, since these two planets host the largest 
number of regular satellites and have enjoyed a long history of detailed 
observations. In this framework, the primary challenge is to understand the 
differences between the number of major moons.

One of the leading formation models comes from a series of papers developed
by \cite{canup:2002,canup:2006,canup:2009} and is known as the actively 
supplied gaseous accretion disk model. In this model, dust grains within a 
circumplanetary disk stick and grow to form satellitesimals, which then migrate 
via type I migration and are disposed within $\sim10^5$\,yrs \cite{tanaka:2002}. 
Continuous mass-infall form the protoplanetary disk maintains a peak 
circumplanetary disk density of $\sim100$\,g\,cm$^{-2}$, allowing new 
satellitesimals to continuously grow. Once the planet has opened up a gap in the 
protoplanetary disk, the active supply halts and the circumplanetary disk 
rapidly diffuses in $\sim10^3$\,yrs, thus ``freezing'' the remaining satellites 
in place. The final total satellite mass is therefore given by a balance between 
type I migration disposal and the repeated satellitesimal accretion rate, which 
\cite{canup:2002} argue to be universally of order 
$(\sum M_S)/M_P\sim10^{-4}$. In this model, the Jovian/Kronian differences are 
proposed to be due to presence of an inner cavity within the circum-Jovian disk, 
which Saturn was unable to form due its distinct mass and semi-major axis 
\cite{sasaki:2010}.

An alternative model comes from \cite{mosqueira:2003a,mosqueira:2003b}
and \cite{estrada:2009}: the solids enhanced minimum mass 
model. Here, the planet hosts a two-component disk comprised of a dense inner 
sub-disk surrounded by a lower density outer disk. This results in a much
longer satellite migration timescale than the associated formation timescale. The 
model qualitatively describes the expected mass ratios, but does not provide 
definitive predictions-- unlike the actively supplied disk accretion model.

\subsection{Formation of Irregular Satellites}

In this Review, I consider irregular satellites to be those which initially 
formed not as a satellite to the planet in question, but somehow ended up so via 
a capture or impact event. Before this event, the satellite may have been a 
planet/dwarf planet itself, a Trojan, a satellite of another planet or a binary 
pair. For moons to form via a capture or impact, one clearly requires strong 
dynamical mixing leading to close encounters. The ultimate outcome of an 
encounter will chiefly depend upon how close these encounters are (the impact 
parameter) and the relative masses and velocities of the bodies involved. In 
this sense, one can consider impacts to be a special case of a capture, where 
the impact parameter of the event is less than one planetary radius.

Giant planets must frequently migrate given the observed population of 
hot Jupiters, which certainly could not have formed in-situ 
\cite{lin:1996}. During this inward migration, the giant planet may encounter 
terrestrial planets and in such encounters a capture could transpire. In all 
capture scenarios, one requires the relative velocities of the two bodies to be
below that of the planet's escape velocity, if the putative satellite is to be
captured. This generally requires a deceleration of the putative satellite,
for which several mechanisms have been proposed.

In the case of the Moon, our satellite's composition, rotation and orientation
support the so-called ``giant impact hypothesis'' between the primordial
Earth and a Mars-sized body some 4.5\,Gyr ago \cite{hartman:1975}.
Whilst this hypothesis explains much of the available evidence, it is unclear 
how frequently such events may occur or how special the Moon may be. Indeed, 
this question strongly motivates seeking observational evidence for exomoons.

A mechanism for producing irregular moons, requiring a less finely tuned 
impact parameter, comes from the ``binary-exchange mechanism''
\cite{agnor:2006}. Here, a binary of similar masses
encounters a giant planet, causing tidal disruption of the binary. If the
binary has the correct sense of revolution, then the member closest to the
giant planet can be strongly decelerated by virtue of the binary's interior
motion. This body is then captured to an initially highly inclined and/or 
eccentric orbit, whilst the other member of the binary is ejected. This 
hypothesis has gained favor for the origin of Triton and \cite{williams:2013} 
have suggested that for exomoons, mass ratios up to 10:1 may result.

Once a moon has been initially captured, the dynamically hot orbit is rapidly
circularized via tides. \cite{porter:2011} estimate this process
occurs on a timescale of $10^3$\,yrs leading to 25-60\% of satellites surviving
into circularized, stable orbits. Irregular satellites therefore offer an
apparently plausible mechanism for the origin of large moons, such as the
Moon and Triton. 

\subsection{Likely Formation of Detectable Exomoons}

Current theories of satellite formation therefore indicate that the
highest satellite-to-planet mass ratio moons are more feasibly formed via a 
capture/impact event, rather than precipitating out of the circumplanetary disk. 
This point is crucial since current observational searches for exomoons are only 
sensitive to high mass ratios exomoons of order of 1\% to 10\%, as will be 
discussed later. As seen in the next subsection, an irregular moon can in fact 
have a longer survival time in the subsequent orbital evolution too. On this 
basis, current models imply that any exomoons large enough to be detected in the 
near future are more likely to be irregular moons.

A large mass-ratio exomoon captured around an exoplanet is presumably a
relatively rare event. The Moon-Earth system presents the only known example of
a $\mathcal{O}[10^{-2}]$ mass-ratio in the Solar System (ignoring the dwarf
planets) leading to the conclusion that the conditions under which the Moon
formed were certainly not typical for all endoplanets. Determining 
the occurence rate of such large mass ratio exomoons will of course be an 
important measurement from observational searches to constrain current theories. 
Producing these events requires either a very low impact parameter or a low 
velocity encounter between two bodies \cite{williams:2013}, for which either
scenario seems to be the tail-end of the probability distribution of
possible outcomes. With these arguments in mind, it would seem that likely that
in the apparently rare instances where a moon has a $\mathcal{O}[10^{-2}]$ 
mass-ratio to the host planet, it is likely the only significant exomoon in 
orbit of said planet. For reasons discussed in the last subsection, I here posit 
that any exomoon detected in the near future will not co-exist with any other 
major satellites. Accordingly, moon-moon interactions will not significantly 
feature in the subsequent orbital evolution.

\subsection{Long-Term Evolution of Exomoons}

For an irregular satellite, once the initial rapid tidal circularization has 
completed in just a few thousands years, the subsequent orbital evolution of the 
moon is dominated by tidal interactions with the host planet and gravitational 
interactions with other satellites in the system. For reasons discussed in the
last subsection, I here consider that any detectable exomoon in the near-future
will not be competing with any other major satellites and thus moon-moon
interactions will not significantly feature in the subsequent orbital evolution.

The only mechanism which therefore contributes significantly to the orbital 
evolution of our exomoon is tidal dissipation. Consider a planet-moon pair with 
a near-circular orbit and rotation periods shorter than the revolution period. 
Here, the tidal evolution occurs in three stages. Firstly, the torque applied by 
the planet's gravity on bulges induced on the moon leads to tidal 
friction and a gradual decrease in the rotation rate of the satellite, 
ultimately leading to tidal locking of the satellite. By conservation of angular 
momentum, the orbit of the satellite is slightly raised by this process. Next,
the same tidal torque slows down the rotation of the planet over a longer
timescale and again lifts the orbit of the satellite. Finally, the planet's
rotation rate becomes synchronized with the revolution rate so both bodies are
tidally locked into a spin-orbit resonance and after this point tidal
evolution ceases. For the Earth-Moon system this is some 50\,Gyr away and would
occur at a period of 47\,d. In practice, a slight drag force on the satellite
would decay the orbit until the tidal evolution kicks in again, but now in 
reverse, causing inward migration, since now the position of the tidal bulge 
trails the planet-moon axis.

\cite{barnes:2002} modeled this process for a variety of
planet-moon pairs and found that even for habitable-zone planets, the Hill
sphere (the volume of stable moon orbits) is large enough that losing a moon
through tidal evolution takes many Gyr. \cite{domingos:2006} argue
the maximum orbital radius for stable moon orbits is actually about twice as
large for retrograde satellites, suggesting they should have much longer
tidal lifetimes than their prograde counterparts. In conclusion, the process of 
tidal evolution for a planet-moon pair seems unlikely to be a significant 
bottleneck to a present day population of large exomoons. In contrast, the
initial formation of such moons seems to require some fine tuning. Therefore,
the very presence of a large exomoon would be interesting from a formation
perspective while the present-day orbital parameters would be interesting from
an evolution perspective.

\subsection{Stripping a Planet of its Moons}

For a single star with just one circular orbit planet which harbors a single 
moon, orbital evolution should indeed proceed as described above. In reality, 
additional bodies and eccentricity can evolve the planet's orbit and lead to 
scattering events. If a planet with a moon migrates inwards through the 
protoplanetary disk in the first few million years after formation, the Hill 
sphere of the planet will shrink much more rapidly than the timescale for 
orbital evolution of the moons. This causes any moons to essentially find 
themselves suddenly outside the Hill sphere and thus ejected, as discussed by 
\cite{namouni:2010}. The author suggests giant planets which have migrated 
interior to 0.1\,AU will have lost most of their initial moons by this process.
For this reason, observational searches should likely avoid hot-Jupiters 
systems.

\cite{gong:2013} recently discuss how solar systems with just a
single planetary member may imply a history of severe planet-planet
scattering. Close encounters between planets is found to efficiently strip
the planets of their moons, although in some cases exchanges of satellites
may occur. For this reason, \cite{gong:2013} recommend observational
searches focus on multi-planet systems. In practice, it is practically 
impossible to determine that a star absolutely only hosts one planet,
especially for transiting systems, and so this advice should be taken with
some caution.

\section{DETECTION TECHNIQUES}

\subsection{Overview}

Numerous techniques for detecting exomoons have been proposed since
their planetary counterparts began to be found. Here, I will focus exclusively 
on those detections methods which rely on the transit technique - that 
is, when an exoplanet periodically passes in front of the host star causing a
repeated decrease in the apparent brightness. Transits only occur for planets 
which happen to have nearly edge-on orbits relative to our line of sight. 
For a circular orbit planet, the geometric probability is simply the radius of
the star divided by the semi-major axis and is $\mathcal{O}[10^{-2}]$.
Despite this drawback, the transit method has emerged has one of
the dominant techniques for discovering exoplanets and arguably \textit{the}
dominant technique for exoplanet characterization. 

The reason why the transit method has enjoyed so much focus, funding and
interest from the community is perhaps down to the simplicity of the observation 
and the rich plethora of planet properties which can inferred. As an example, 
the shape of the transit light curve encodes the inclination of the orbit 
\cite{seager:2003}, which can be combined with a radial 
velocity (RV) signal to back out the planet's true mass (rather than just the 
\textit{minimum} mass, which RVs usually provide). Characterizing the 
atmospheres of exoplanets has sprung forth an entire field of scientific 
activity, usually by seeking changes in the depths of transits in different 
wavelengths, betraying the presence of absorbing molecules in the atmospheres 
\cite{seager:2000}.

Eclipsing bodies have a long history of providing deep physical insights
and sensitivies and the cost of fewer objects to study \cite{russell:1948}.
It is for similar reasons that the transit method provides viable methods
for detecting exomoons. However, one should proceed under no illusion that this
is by any means easy, even when leveraging the unique insights provides by
transits.

Other methods for detecting exomoons include pulsar timing \cite{lewis:2008}, 
which is limited to looking at a relatively unusual type of
star. Whilst a direct image of an exomoon is some way off given their likely
tiny size, it has been argued that evidence for circumplanetary disks has been
captured in exoplanet direct imaging \cite{kalas:2008}. The 
microlensing technique is also capable of finding exomoons and has even
reported a possible candidate recently, around a free-floating planet 
\cite{bennett:2014}. However in general, and certainly in the case of 
this candidate, it is not possible to uniquely determine the mass of the lens
pair, meaning that a low-mass star with a planet is also a viable explanation to 
the observations \cite{bennett:2014}. Additionally, microlensing 
events are a-periodic and thus there is no prospect of future observations to 
confirm any claimed detections.

In conclusion, I would argue that the transit method is the favored technique 
for discovering exomoons, allowing one to i] see repeated events and 
make causal predictions regarding future observations (a fundamental pillar
of the scientific method) ii] detect exomoons around any luminous star which
has a transiting planet iii] characterize any discoveries with future 
observations to measure the atmosphere and orbit iv] determine all 
of the basic physical parameters of the moon from time series photometry
alone. In the following subsections, I will therefore focus exclusively on the
transit method and discuss how the claimed feats mentioned above are possible.

For transiting systems, there are two broad class of observable effects due to
an exomoon: dynamical effects and eclipse effects. I will discuss each of these
in the following subsections.

\subsection{Dynamical Effects} 

\subsubsection{Transit Timing Variations (TTV)}

The first scientific paper to propose a method for detecting the moons of 
exoplanets suggested seeking variations in the times of the transit events 
\cite{sartoretti:1999}. Further, this paper appears to be the
first usage of the term ``transit timing variations'' (TTV), which became
lately more commonly associated with detecting planet-planet gravitational
interactions \cite{agol:2005,holman:2005}. As
shown in Figure~\ref{fig:ttv}, a planet with a moon will exhibit reflex 
motion in response to the companion's mass on top of the usual Keplerian
orbit around the host star. This reflex motion causes changes in the position
and velocity of the planet relative to a strict Keplerian, and these changes
manifest as transit timing variations (TTV) and transit duration variations
(TDV-V) respectively. It is worth noting that conceptually TTV is similar to the 
asterometric method of finding planets, whereas TDV-V is similar to that of 
the radial velocity method.
Whilst \cite{sartoretti:1999} derived the expected TTV 
amplitude for circular orbits, the more general eccentric case was found by
\cite{kipping:2011} to be given by:

\begin{figure}[ht]
\centering
\includegraphics[width = 0.5\textwidth]{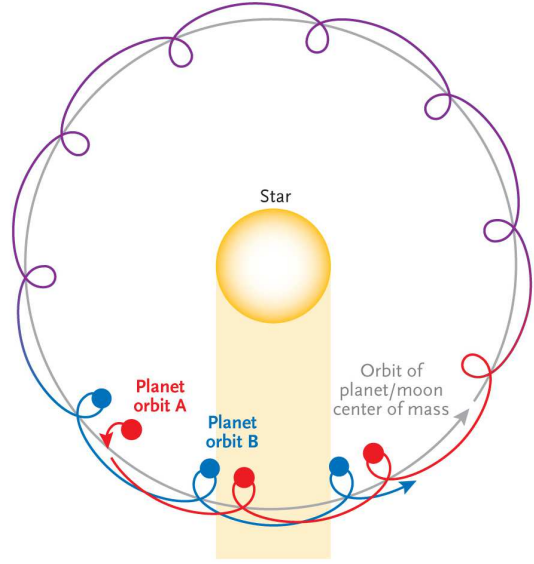}
\caption{Schematic illustrating the (exaggerated) motion of a planet around its 
host star in response to a massive companion satellite (not shown). The planet's 
position and velocity deviate from that expected from a simple Keplerian (gray 
line), producing variations in the times and durations of the transits, 
respectively.
}
\label{fig:ttv}
\end{figure}

\begin{align}
\delta_{\mathrm{TTV}} &= \frac{a_S M_S}{a_P M_P} P_P \frac{(1-e_S^2) \sqrt{1-e_P^2}}{(1+e_P\sin\omega_P)} \sqrt{\frac{\Phi_{\mathrm{TTV}}}{2\pi}},
\label{eqn:TTVrms}
\end{align}

where $\delta_{\mathrm{TTV}}$ is the root-mean-square (RMS) TTV amplitude, 
$a_S$ and $a_P$ are the planet-satellite and planet-star semi-major axes,
$M_S$ and $M_P$ are the satellite and planet masses, $e_S$ and $e_P$ are the 
orbital eccentricities of the satellite and the planet, $\omega_P$ is the 
argument of periastron of the planet and $\Phi_{\mathrm{TTV}}$ is absorbs many
of the effects of orbital eccentricity and is defined in \cite{kipping:2011}. 
The effect is typically of order of seconds to minutes for terrestrial
moons around gas giant planets, which is comparable to \emph{Kepler's} timing
precision.

From Equation~\ref{eqn:TTVrms} and ignoring eccentricity, one can see that the 
amplitude scales as $\sim M_S a_S$ i.e. $M_S$ and $a_S$ are fully degenerate. 
In principle, one could determine the period of the satellite, $P_S$, by the 
periodicity of the TTV signal, which could then be converted to $a_S$ via 
Kepler's Third Law. However in practice this is not possible, since the moon's 
period must always satisfy $P_S \leq P_P/\sqrt{3}$ \cite{kipping:2009a} for 
dynamical stability, which is below the Nyquist frequency. Therefore, a unique 
solution for $P_S$, and thus $a_S$, is not possible. Crucially then, the TTV 
RMS amplitude, which is simply found by measuring the scatter in
the data, is all we can infer and this displays a complete degeneracy between 
the moon's mass and orbital separation.

This problem with using TTVs to detect exomoons is compounded by the fact
planet-planet interactions can also induce confounding TTV signatures
\cite{agol:2005,holman:2005}, which would be a 
source of false-positives. For these reasons, TTV is generally unable to
uniquely detect an exomoon. This motivated the community to develop alternative
or complementary techniques to solve these issues.

\subsubsection{Velocity induced Transit Duration Variations (TDV-V)}

As discussed earlier, the reflex motion of the planet, in response to the moon,
causes changes in the velocity of the planet, as well as the position. Velocity 
changes lead to variations in the durations of the transits
\cite{kipping:2009a,kipping:2009b}, which can present a detectable signal; 
so-called velocity-induced transit duration variations (TDV-Vs). The 
RMS amplitude of the TDV signal is given by \cite{kipping:2011}:

\begin{align}
\delta_{\mathrm{TDV-V}} &= \tilde{T}_B \Bigg(\frac{a_S M_S P_P}{a_P M_P P_S}\Bigg) \Bigg(\frac{\sqrt{1-e_P^2} \sqrt{\Phi_{\mathrm{TDV-V}}/2\pi} }{\sqrt{1-e_S^2} (1+e_P\sin\omega_P)}\Bigg),
\label{eqn:TDVrms}
\end{align}

where $\tilde{T}_B$ is the mean transit duration and $\Phi_{\mathrm{TDV-V}}$
absorbs several effects due to orbital eccentricity and is defined in
\cite{kipping:2011}. The effect is typically of order of seconds to minutes 
for terrestrial moons around gas giant planets, an amplitude similar to that of 
TTVs. Inspection of Equation~\ref{eqn:TDVrms} reveals that the TDV-V amplitude 
scales as $\sim M_S a_S P_S^{-1}$. Recall that the TTV amplitude 
scales as $\sim M_S a_S$ though, meaning that $P_S$ (and thus $a_S$ via Kepler's 
Third Law) can be determined simply by comparing the TDV-V RMS amplitude to the 
TTV RMS amplitude (for circular orbits). This solves the undersampling issue 
encountered using TTVs in isolation.

An additional benefit of TDV-Vs, is that the planet-planet interactions do
not usually produce detectable TDVs, and when they do the signal is expected to 
be either perfectly in phase or in anti-phase ($\Delta\phi=0$ or $\pi$), as seen
in \cite{nesvorny:2013}. In contrast, the reflex motion of the 
planet is essentially two-dimensional simple harmonic motion and thus the 
position (TTVs) and velocity (TDV-Vs) exhibit a $\Delta\phi=\pi/2$ phase shift
\cite{kipping:2009a}. This specific signature enables the unique 
identification of exomoons. Despite the signals being undersampled, this
phase shift can be inferred by cross-correlating the normalized squared
TTVs and TDV-Vs \cite{awiphan:2013}.

\begin{figure*}[ht]
\centering
\includegraphics[width = 0.9\textwidth]{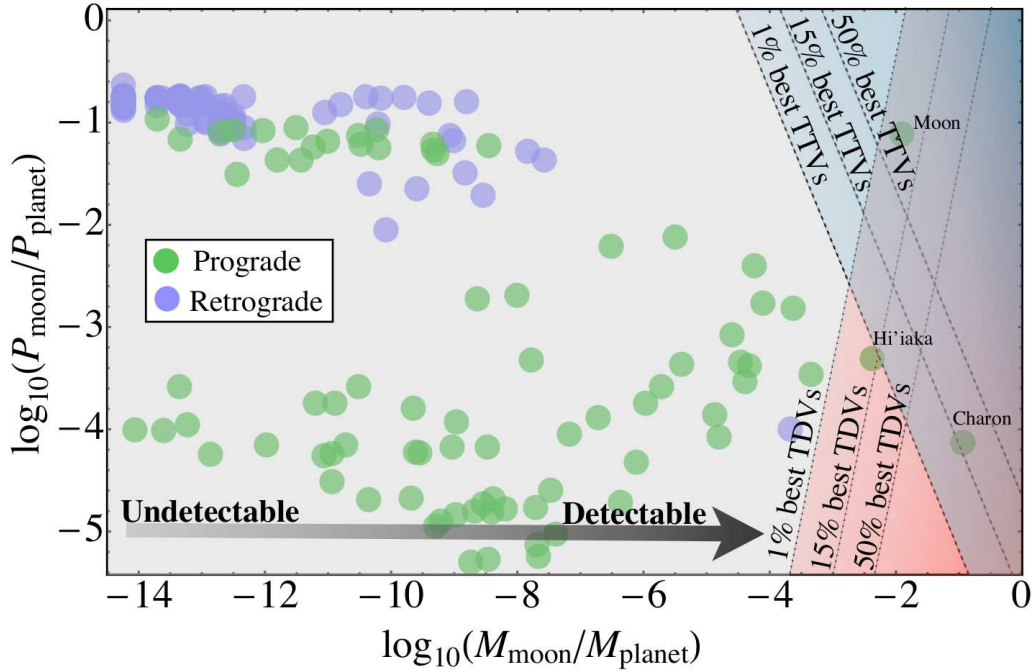}
\caption{Period-ratio versus mass-ratio scatter plot of the Solar System moons. 
Transit timing and duration variations (TTV and TDV) exhibit complementary 
sensitivities with the period-ratio. Using the \emph{Kepler} timing measurements 
from \cite{ford:2012}, one can see that the tip of the observed 
distribution is detectable. The above assumes a planetary period of 100\,d and a 
baseline of 4.35\,years of Kepler data.}
\label{fig:solar}
\end{figure*}

Despite the mathematical elegance of combining TTVs and TDV-Vs, one important
limitation is that we require the detection of \textit{both} effects, which is
in practice very challenging. One can easily see this by virtue of the amplitude
scalings. Specifically, TTVs scale as $(M_S a_S)$ where as TDV-Vs scale as
$(M_S a_S^{-1/2})$. Therefore, moons close to the planet will cause a large TDV
but a much smaller TTV, and vice versa. This means that in the absence of
infinite signal-to-noise, only intermediate separation moons can be detected.
\cite{kipping:2009} explored this issue in detail simulating a 
variety of planet-moon configurations and concluded that around one sixth of 
\emph{Kepler's} target stars would provide sufficient signal-to-noise to detect
both effects for Earth-mass moons with a 25\% recovery rate. This calculation
assumes various priors on the orbital elements for exomoons though, which one
generally does not know since no Earth-mass moons exist in the Solar System.
However, one can take the properties of the known Solar System moons and
simulate their detectability using \emph{Kepler}. To do so, one must move the 
orbits of the outer planets inwards, since the \emph{Kepler} baseline only
covers 4.35\,years. Opting for a planetary period of 100\,d then and using
the typical \emph{Kepler} timing uncertainties reported in \cite{ford:2012}, one 
can see in Figure~\ref{fig:solar} that several endomoons would be
detectable.

\subsubsection{Transit Impact Parameter induced Transit Duration Variations 
(TDV-TIP)}

I briefly discuss an additional second-order TDV effect known as
transit impact parameter induced transit duration variations, TDV-TIP
\cite{kipping:2009b}. So far, the TTV and TDV-V effects have ignored motion
occurring out-of-the-plane. In reality, orbits exist in three-dimensions and
the same positional variations which gives rise to translational shifts
(leading to TTVs) will also have some non-zero component in the orthogonal
direction. Inclination in either (or both) the planet and the satellite
relative to the line-of-sight lead to the reflex motion slightly adjusting
the ``height'' of the transit chord, or more specifically the transit
impact parameter. Since the transit impact parameter strongly affects the 
transit duration \cite{seager:2003}, then TDVs ensue.

TDV-TIP is typically an order of magnitude lower amplitude than TDV-V, for
nearly coplanar orbits \cite{kipping:2009b}, at the level of sub-seconds
to seconds. In extremely inclined orbits, the signal can become dominant
though and should always be accounted for when modeling timing signals. As 
proved in \cite{kipping:2009b}, their detection would allow for the 
inference of the sense of orbital motion of the moon i.e. whether it is prograde 
or retrograde. This determination would have important ramifications for
understanding the origin of said exomoon, since regular moons do not form in
retrograde orbits. It becomes clear, then, that the transit light curve of a 
planet-moon system contains information on almost all of the orbital elements 
and thus once a system is identified, detailed follow-up could characterize 
these subtle effects.

\subsection{Eclipse Effects} 

Although the direct transits of an exomoon may seem the most obvious way to
detect such objects, efforts to model these events trailed behind the dynamical
methods by a few years. Whereas the dynamical effects provide the mass of
the satellite, eclipse effects would reveal the radius of the moon. Combining 
the two could then reveal the density of the satellite and thus
inferences about the composition.

Eclipse effects come in two flavors; ``auxiliary transits'' and 
``mutual events''. The former are simply those cases where the projected
disk of the moon does not overlap with that of the planet but does overlap with 
that of the star. The latter is where all three disks overlap. These two
effects have very different morphologies and modeling approaches.

\subsubsection{Auxiliary Transits}

In an auxiliary transit, the disk of the moon does not overlap with that of
the star. This means one can simply compute the transit light curve profile
caused by the planet and star separately add combine them together via:

\begin{align}
I(t) = 1 - \Big[ (1-I_P(t)) + (1-I_S(t)) \Big],
\end{align}

where $I(t)$ is the overall normalized intensity of the star as a function of 
time, whereas $I_P(t)$ and $I_S(t)$ are the components due to the planet and
satellite in isolation. These individual components may be simply computed
using the standard \cite{mandel:2002}, for example. \cite{simon:2012}
suggest that a planet with a very short-period could be folded to reveal
excess scatter either side of the main transit event, due to auxiliary transits.
Whilst simply seeking scatter would be a rapid way of detecting exomoons, it
does not address the uniqueness of such an interpretation nor is it likely that
short-period planets would exomoons anyway due to their much smaller Hill
sphere.

\subsubsection{Mutual Events}

A mutual event is where all three disks overlap, meaning that one must compute 
the area of common overlap between the disks. This may be achieved using the
analytic solutions from \cite{fewell:2006}. Typically, these events would
occur by the planet and moon entering transit together with a small
projected separation. The moon then skips across the face of the planet whilst
both of the disks still overlap the star. During this time, the apparent
flux of the star seems to \textit{increase}, since one goes from having the 
planet and moon blocking the star to just the planet (since now the moon is 
hidden in front of or behind the planet). The moon then appears on the opposite 
side of the planet with a small separation and the two disks exit transit.

\begin{figure*}[ht]
\centering
\includegraphics[width = 0.9\textwidth]{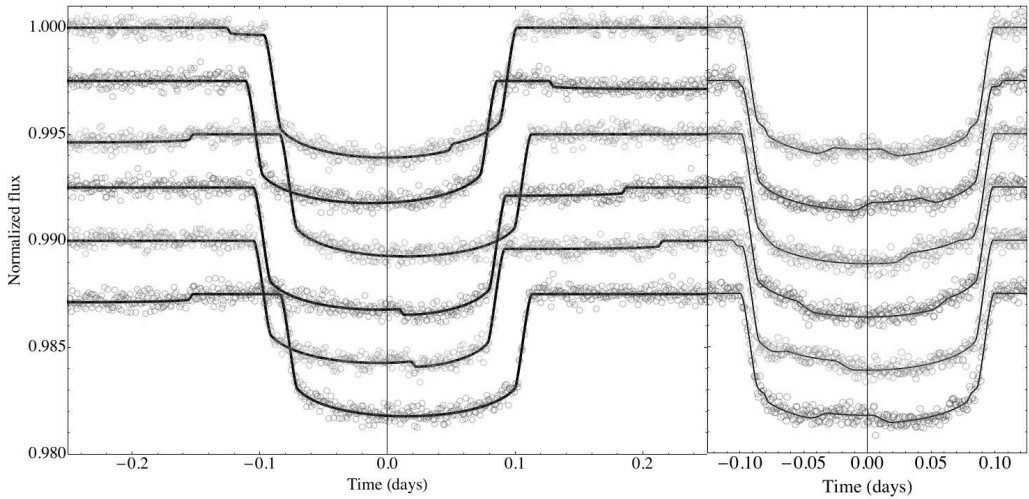}
\caption{\emph{(Left)} Six simulated transits using \luna\ \cite{luna:2011} 
of a HZ Neptune around an M2 star with an Earth-like moon on a wide orbit 
(90\% of the Hill radius). The moon can be seen to exhibit auxiliary transits 
and induce TTVs. \emph{(Right)} Same as left, except the moon is now on a 
close-in orbit (5\% of the Hill radius), causing mutual events. Both plots show 
typical \emph{Kepler} noise properties for a $12^{\mathrm{th}}$-magnitude star 
observed in short-cadence.}
\label{fig:luna}
\end{figure*}

Examples of mutual events are shown in Figure~\ref{fig:luna}. These events
can look very similar to that of a planet passing over a dark starspot, leading 
to a potential source of false positive (e.g. see \cite{rabus:2009}).
Mutual events can also, in some very rare instances, occur between two
planets orbiting the same star \cite{hirano:2012}. Finally, it is
worth noting that mutual events require a closely-aligned system for such
events to be possible and a short-period/close-separation for such events
to be frequent enough to expect to observe.

\subsection{Photodynamical Algorithms} 

A photodynamical algorithm aims to model both the eclipse effects (``photo'')
and the dynamical effects simultaneously. Such codes have been leveraged
in studying planet-planet interactions \cite{lissauer:2011} and 
circumplanetary planets \cite{carter:2011}. For exomoons,
photodynamical algorithms don't merely serve to provide a better estimate
of the system parameters but also fundamentally push one's sensitivity to
small exomoons to the limit. This is because, in principle, a photodynamical
algorithm models all of the observational imprints of an exomoon. Consequently,
even if the significance of the TTVs, TDV-Vs, TDV-TIPs, auxiliary transits and
mutual events are independently quite low, the combination of all these can 
push the significance above the discovery threshold. Since the field current
lacks a confirmed exomoon detection, this point in fact represents the primary 
goal of the current generation of photodynamical algorithms - to enhance the act 
of detection.

With this goal in mind, practical considerations will also feature in the
design of any such algorithm. For example, search techniques typically attempt
billions of different possible star-planet-moon configurations and
compare them to the data, in order to identify if the observations are well
explained by including a moon or not. This is generally necessary due to
multimodal parameter space featuring complex and curved degeneracies 
\cite{kipping:2012}, i.e. simple downhill algorithms are 
nearly useless for seeking evidence of exomoons. This requirement shapes the
design of any algorithm since the model needs to be extremely fast to execute
in order to make it feasible to scan multiple data sets for exomoons.
Consequently, analytic photodynamical algorithms are the only
practical solution.

Several approaches to producing a photodyanmical algorithm, to various degrees
of completion, are available in the literature. \cite{szabo:2006} 
suggested re-defining the classic definition of the transit mid-time to a flux 
weighted ``photo-center'', which would incorporate some of the effects of both 
eclipses and dynamics. However, this approach does not account for orbital
eccentricity or any other dynamical effects except TTV. \cite{sato:2009,
sato:2010} and \cite{tusnski:2011} provide eclipse models which 
ignore the dynamical effects but reproduce the ``photo'' effects. Naturally,
without a mass estimate of the moon it is unclear how reasonably one
could uniquely identify signals as being due to an exomoon.

The only published full analytic photodynamical algorithm for exomoons, at the 
time of writing, is that of \luna\ \cite{luna:2011}. This code 
dynamically progress the position of the planet and moon at every time stamp, 
using a restricted three-body solution, and reproduces limb darkened light 
curves simulating both auxiliary and mutual transits. Simulations from \luna\ 
are shown in Figure~\ref{fig:luna}, illustrating the typical light curve 
morphologies of both auxiliary and mutual transits.

\section{RESULTS TO DATE}

In this section, I will focus on results published by the ``Hunt for Exomoons
with Kepler'' (HEK) project \cite{kipping:2012}. To date, this is
the only search project in the literature. HEK seeks to determine the occurrence
rate of large moons around viable planet hosts, $\eta_{\leftmoon}$. 

The search is conducted using the photodynamical algorithm, \luna, coupled with 
a multimodal nested algorithm called \multi\ \cite{feroz:2008,feroz:2009}. 
\multi\ is not only exhaustive, identifying all
of the modes and complex curved degeneracies, but also provides the Bayesian
evidence, $\mathcal{Z}$, of any model attempted. When one takes the ratio
of $\mathcal{Z}$ between two models, say a planet-only versus a planet-with-moon 
model, the result is equal to the odds ratio between the models. In other words, 
we can determine which model is the most likely one. The Bayesian evidence 
penalizes models for using more parameters and thus includes a built-in Occam's 
razor. This is crucial when seeking exomoons, since including what is 
essentially a bunch of perturbation terms will always yield a better likelihood 
due to the higher number of degrees of freedom.

These Bayesian fits and parameters searches are computationally expensive 
though, taking up to 50\,years of CPU time for a single system 
\cite{kipping:2013b}. For this reason, HEK has been limited in the number of 
targets which can be investigated, focussing efforts on those planets which are 
thought to be capable of both maintaing moons via the tidal theory of 
\cite{barnes:2002} and presenting a detectable signal via the expected timing
effects \cite{kipping:2009a,kipping:2009b}.

\bigskip
\noindent
\textbf{4.1 Constraints for 17 Exoplanets} 
\bigskip

\begin{figure*}[ht]
\centering
\includegraphics[width = 1.0\textwidth]{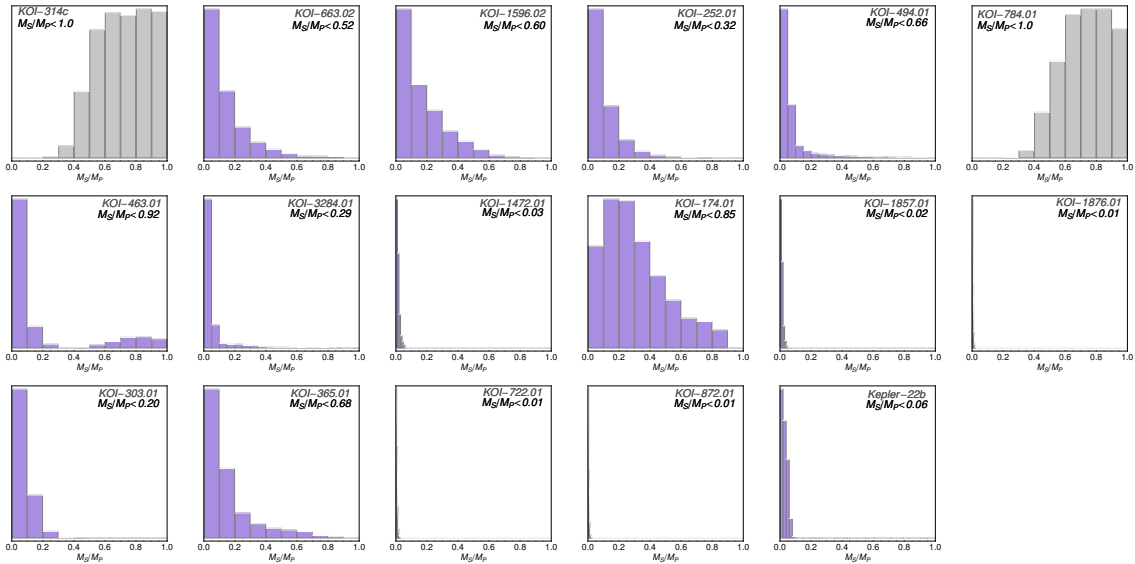}
\caption{Empirical posterior distributions for the mass ratio between a
putative satellite and 17 \emph{Kepler} planetary candidates surveyed by
the HEK project. 95\% confidence upper limits provided in the top-right.
KOI-314c and KOI-784.01 are grayed out as planet-planet interactions lead to
spurious detections.
}
\label{fig:Msp}
\end{figure*}

Spanning the papers \cite{nesvorny:2012,kipping:2013a,kipping:2013b,
kipping:2014}, a total of 17 \emph{Kepler} planetary candidates have been
surveyed by the HEK project, to date. There are no confirmed exomoons 
discoveries but upper limits on the mass ratio between a putative satellite
and the host planet, $(M_S/M_P)$, in all cases, which are shown in 
Figure~\ref{fig:Msp}.

Although the light curve fits only reveal $(M_S/M_P)$ directly, one can
translate these values into physical masses by invoking a mass-radius relation
for the planet. Using the \cite{weiss:2014} relation, six of the
constraints correspond to sub-Earth mass sensitivity. \cite{kipping:2013b}
went further in the case of Kepler-22b by injecting a fake Earth-like
moon into the observational data and retrieving it to $>8$\,$\sigma$ confidence,
demonstrating that HEK is certainly able to detect Earth-like moons should
they exist. The absence of confirmed discoveries is therefore speaking to the
underlying occurrence rate. At this point, the sample is clearly too small
to derive a robust $\eta_{\leftmoon}$ estimate, but this remains the ultimate
goal of the HEK project.

\subsection{Ancillary Science} 

Whilst the HEK project has probed down to sub-Earth masses and
Earth-Moon mass ratios in its quest for exomoons, several interesting
ancillary science discoveries have emerged from the project too. For example,
\cite{nesvorny:2012} reported the first non-transiting planet
discovered (with a unique orbit) via the TTV method in the KOI-872 system. This 
opens the door to not only discovering a new subset of planets hidden in the 
\emph{Kepler} data, but also measuring precise masses and orbital elements in 
these dynamical systems. In this case, the TTVs were initially considered a
plausible moon candidate but dynamical modeling revealed planet-planet
interactions as the only viable model. \cite{nesvorny:2013}
recently repeated the feat for another system, KOI-142, where the TTV amplitude 
was equal to nearly 10\% of the orbital period, a situation the authors dubbed 
as ``The King of TTVs''.

\cite{kipping:2014} recently reported another non-moon TTV signal
for KOI-314, in this case due to interaction between two planets where 
\textit{both} transit their host star. In this case, the outer planet, KOI-314c,
was found to have a mass of $M_P=1.0_{-0.3}^{+0.4}$\,$M_{\oplus}$, making it
the lowest mass transiting planet discovered to date. These cases highlight
the exciting ancillary science resulting from exomoon surveys.

\section{OUTLOOK}

``When will we discover an exomoon?'' is probably the question I get asked 
the most when discussing this subject, yet a definitive prediction cannot be 
made for the fundamental reason that the occurrence rate of large moons is 
presently unknown. Nevertheless, I consider here the future outlook and
prospects for success.

The \emph{Kepler} data undoubtedly presents the only viable channel for 
detecting transiting exomoons with past or present observatories. The 4+\,year
baseline of continuous, high-precision photometry makes it greatly superior to 
\emph{CoRoT}, HST or MOST data collected to date. The lower temporal baseline
of the planned TESS (launch 2017-2018) mission makes it less attractive than 
\emph{Kepler} for detecting exomoons, but PLATO 2.0 (launch 2024) may offer
opportunities with a planned 2\,year baseline at \emph{Kepler}-like precision.
JWST will also be too competitive to acquire the large temporal baselines
necessary. It therefore appears that if \emph{Kepler} is unable to succeed in 
discovering an exomoon, future missions (up to the mid-2020's) will be generally 
less suitable.

With this point in mind, what are the prospects of \emph{Kepler} succeeding?
The failure of a second reaction wheel ended the nominal \emph{Kepler} mission.
A planned K2 mission will exploit a two-wheel \emph{Kepler}, but since
it uses radiation pressure as a third torque, pointing is limited to near the
ecliptic plane, making a re-observation of the original \emph{Kepler} field
impossible. In this sense, the 4.35\,years of archival \emph{Kepler} data
represent our best hope for a discovery. Whilst further data would of course
benefit the exomoon hunt, 4.35\,years is sufficient to identify more than 
half-a-dozen transits of planets with periods of 8\,months or less, which
should be sufficient to identify a large moon.

The HEK project has surveyed just 17 targets to date, yet $\sim250$ planetary 
candidates are estimated to have the correct properties to maintain a detectable 
exomoon. In general, issues with planet false-positives, time-correlated noise 
and stellar activity limit the feasible sample to a number less than $250$. 
Despite this, at least 100 targets should provide the required data quality to 
seek exomoons and the HEK project intends to survey all of these over the next 
2-3\,years.

What kind of sensitivity level should we expect these ``good'' targets to yield?
The light curve fits directly reveal $(M_S/M_P)$ and the typical constraints
obtained to date (Figure~\ref{fig:Msp}) vary considerably from system to system, 
depending upon the transit profile shape, orbital period, noise properties of 
the data and presence of perturbing planets in the system. Although only 
17 objects have been surveyed, one can use this sample to estimate the 
approximate sensivity of the HEK survey. 

\begin{figure}[ht]
\centering
\includegraphics[width = 0.5\textwidth]{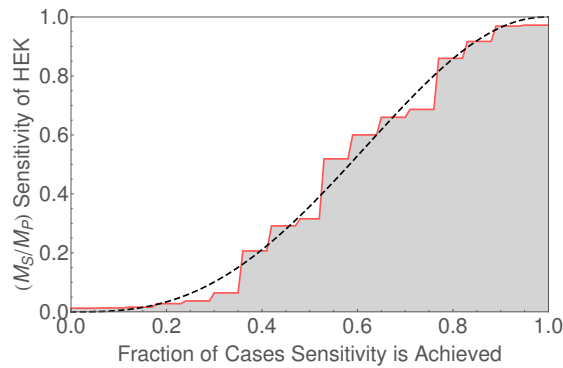}
\caption{Empirical sensitivity of the HEK project to the satellite planet mass 
ratio (red line), based on 17 published results. HEK is able to detect 
Earth-Moon like mass ratios in $\sim14$\% of cases. Black line shows a smoothed
functional fit.
}
\label{fig:sensitivity}
\end{figure}

Figure~\ref{fig:sensitivity} shows the mass-ratio sensitivity as a 
function of the fraction of cases in which the HEK project achieves the said
sensitivity. Using the very flexible cumulative density Beta function provides a 
reasonable smooth fit through the results to date. This function implies
HEK is able to achieve the Earth:Moon mass ratio ($=0.012$) in $\sim14$\% of 
cases and a Pluto:Charon mass ratio of ($=0.116$) in $\sim32$\% of cases. If 
such quasi-binary planet-moon systems are common, then HEK should expect to find
them after sampling $\mathcal{O}[10^2]$ systems. 

In conclusion, whether HEK discovers several moons or none at all, the 
results will reveal just how unique the Solar System is. If large moons are
common, then temperate satellites with atmospheres and perhaps even life may
be a frequent occurrence. If large moons are exceptionally rare, then life in
the Universe may be mostly bound to planets and even Earth-Moon quasi-binaries 
would be a rarity, which may have important implications for the uniqueness of
our home world. Either way, though it pushes the instrument to its very limits,
\emph{Kepler} can deliver the answer. It may take more time and analysis than
discovering planets, but the occurrence rate of exomoons lies buried within the 
\emph{Kepler} data. Its beyond-the-grave determination will shape how we think
about the uniqueness of own home and our place in the Universe.


\clearpage
\begin{thebibliography}{99}
\bibitem{campbell:1988}
Campbell B., Walker G. A. H. \& Yang. S. 1988, \apj, 331, 902
\bibitem{latham:1989}
Latham D. W. et al. 1989, \nature, 339, 38
\bibitem{wolszczan:1992}
Wolszczan, A. \& Frail, D. 1992, \nature, 355, 145
\bibitem{mayor:1995}
Mayor, M. \& Queloz, D. 1995, \nature, 378, 355
\bibitem{marcy:1998}
Marcy, G. W. \& Butler, P. R. 1998, ARA\&A, 36, 57
\bibitem{batalha:2013}
Batalha, N. et al. 2013, \aanda, 204, 24
\bibitem{marcy:2014}
Marcy, G. W. 2014, AAS Meeting \#223, \#91.03
\bibitem{youdin:2013}
Youdin, A. N. \& Kenyon, S. J. 2013, ``From Disks to Planets'' in T. D. Oswalt, 
L. M. French, and P. Kalas, editors, Planets, Stars and Stellar Systems. Volume 
3: Solar and Stellar Planetary Systems.
\bibitem{fressin:2013} 
Fressin, F. et al. 2013, \apj, 766, 81
\bibitem{petigura:2013}
Petigura, E. A., Howard, A. W. \& Marcy, G. W. 2013, \pnas, 110, 19273
\bibitem{rodler:2014}
Rodler, F. \& L\'opez-Morales, M. 2014, \apj, 781, 54
\bibitem{gong:2013}
Gong, Y.-X. et al. 2013, \apj, 769, L14
\bibitem{williams:1997}
Williams, D. M., Kasting, J. F. \& Wade, R. A. 1997, \nature, 385, 234
\bibitem{heller:2012a}
Heller, R. 2012, \aanda, 545, L8
\bibitem{heller:2012b}
Heller, R. \& Barnes, R. 2012, \abio, 13, 18
\bibitem{forgan:2013}
Forgan, D. \& Kipping, D. M. 2013, \mnras, 432, 2994
\bibitem{laskar:1993}
Laskar J., Joutel, F. \& Robutel, P. 1993, \nature, 361, 615
\bibitem{mosqueira:2003a}
Mosqueira, I. \& Estrada, P. R. 2003a, \icarus, 163, 198
\bibitem{canup:2002}
Canup, R. M. \& Ward, W. R. 2002, \aj, 124, 3404
\bibitem{canup:2006}
Canup, R. M. \& Ward, W. R. 2006, \nature, 441, 834
\bibitem{canup:2009}
Canup, R. M. \& Ward, W. R. 2009, in \emph{Europa}, edited by R. T. Pappalardo, 
W. B. McKinnon, and K. Khurana, University of Arizona Press, Tucson, pp. 59
\bibitem{tanaka:2002}
Tanaka, H., Takeuchi, T. \& Ward, W. R. 2002, \apj, 565, 1257
\bibitem{sasaki:2010}
Sasaki, T., Stewart, G. R. \& Ida, S. 2010, \apj, 714, 1052
\bibitem{mosqueira:2003b}
Mosqueira, I. \& Estrada, P. R. 2003b, \icarus, 163, 232
\bibitem{estrada:2009}
Estrada, P. R. et al. 2009, in \emph{Europa}, edited by R. T. Pappalardo, 
W. B. McKinnon, and K. Khurana, University of Arizona Press, Tucson, pp. 27
\bibitem{lin:1996}
Lin, D. N. C., Bodenheimer, P. \& Richardson, D. C. 1996, \nature, 380, 606
\bibitem{hartman:1975}
Hartman, W. K. \& Davis, D. R. 1975, \icarus, 24, 504
\bibitem{agnor:2006}
Agnor, C. B. \& Hamilton, D. P. 2006, \nature, 441, 192
\bibitem{williams:2013}
Williams, D. M. 2013, \abio, 13, 315
\bibitem{porter:2011}
Porter, S. B. \& Grundy, W. M. 2011, \apj, 736, L14
\bibitem{barnes:2002}
Barnes, J. W. \& O’Brien, D. P. 2002, \apj, 575, 1087
\bibitem{domingos:2006}
Domingos, R. C., Winter, O. C. \& Yokoyama, T. 2006, \mnras, 373, 1227
\bibitem{namouni:2010}
Namouni, F. 2010, \apj, 719, L145
\bibitem{seager:2003}
Seager, S. \& Mall\'en-Ornelas, G. 2003, \apj, 585, 1038
\bibitem{seager:2000}
Seager, S. \& Sasselov, D. D. 2000, \apj, 537, 916
\bibitem{russell:1948}
Russell, H. N. 1948, Harvard Coll. Obs. Monograph, 7, 181
\bibitem{lewis:2008}
Lewis, K. M., Sackett, P. S. \& Mardling, R. A. 2008, \aj, 685, L153
\bibitem{kalas:2008}
Kalas, P. et al. 2008, \science, 322, 1345
\bibitem{bennett:2014}
Bennett, D. P. et al. 2014, \apj, submitted
\bibitem{sartoretti:1999}
Sartoretti, P. \& Schneider, J. 1999, \aandas, 134, 553
\bibitem{agol:2005}
Agol, E. et al. 2005, \mnras, 359, 567
\bibitem{holman:2005}
Holman, M. J. \& Murray, N. W. 2005, \science, 307, 1288
\bibitem{kipping:2011}
Kipping D. M. 2011, PhD Thesis, University College London
\bibitem{kipping:2009a}
Kipping D. M. 2009a, \mnras, 392, 181
\bibitem{kipping:2009b}
Kipping D. M. 2009b, \mnras, 396, 1797
\bibitem{nesvorny:2013}
Nesvorn\'y, D. et al. 2013, \apj, 777, 3
\bibitem{awiphan:2013}
Awiphan, S. \& Kerins, E. 2013, \mnras, 432, 2549
\bibitem{kipping:2009}
Kipping D. M., Fossey, S. J. \& Campanella, G. 2009, \mnras, 400, 398
\bibitem{ford:2012}
Ford, E. B. et al. 2012, \apj, 750, 18
\bibitem{mandel:2002}
Mandel, K. \& Agol, E. 2002, \apj, 580, L171
\bibitem{simon:2012}
Simon, A. E. et al. 2012, \mnras, 419, 164
\bibitem{fewell:2006}
Fewell, M. (2006) Tech. Rep. DSTO-TN-0722 (available online at
http://hdl.handle.net/1947/4551).
\bibitem{rabus:2009}
Rabus, M. et al. 2009, \aanda, 494, 391
\bibitem{hirano:2012}
Hirano, T. et al. 2012, \apj, 759, L36
\bibitem{lissauer:2011}
Lissauer, J. J. et al. 2013, \apj, 770, 131
\bibitem{carter:2011}
Carter, J. A. et al. 2011, \science, 331, 562
\bibitem{kipping:2012}
Kipping D. M. et al. 2012, \apj, 750, 115
\bibitem{szabo:2006}
Szab\'o, Gy. M. et al. 2006, \aanda, 450, 395
\bibitem{sato:2009}
Sato, M. \& Asada, H. 2009, \pasj, 61, L29
\bibitem{sato:2010}
Sato, M. \& Asada, H. 2010, \pasj, 62, 1203
\bibitem{tusnski:2011}
Tusnski, L. R. M. \& Valio, A. 2011, \apj, 743, 97
\bibitem{luna:2011}
Kipping D. M. 2011, \mnras, 416, 689
\bibitem{feroz:2008}
Feroz, F. \& Hobson, M. P. 2008, \mnras, 384, 449
\bibitem{feroz:2009}
Feroz, F., Hobson, M. P. \& Bridges, M. 2009, \mnras, 398, 1601
\bibitem{kipping:2013b}
Kipping D. M. et al. 2013b, \apj, 777, 134
\bibitem{nesvorny:2012}
Nesvorn\'y, D. et al. 2012, \science, 336, 1133
\bibitem{kipping:2013a}
Kipping D. M. et al. 2013a, \apj, 770, 101
\bibitem{kipping:2014}
Kipping D. M. et al. 2014, \apj, In Press
\bibitem{weiss:2014}
Weiss, L. M. \& Marcy, G. W. 2014, \apj, 783, L6
\end{thebibliography}
\end{document}